\documentstyle[preprint,aps]{revtex}

\newcommand{\cA}{\mathcal{A}}
\newcommand{\BB}{\mathcal{B}}
\newcommand{\CC}{\mathcal{C}}
\newcommand{\DD}{\mathcal{D}}
\newcommand{\EE}{\mathcal{E}}
\newcommand{\FF}{\mathcal{F}}
\newcommand{\GG}{\mathcal{G}}
\newcommand{\HH}{\mathcal{H}}
\newcommand{\II}{\mathcal{I}}
\newcommand{\JJ}{\mathcal{J}}
\newcommand{\KK}{\mathcal{K}}
\newcommand{\LL}{\mathcal{L}}
\newcommand{\MM}{\mathcal{M}}
\newcommand{\NN}{\mathcal{N}}
\newcommand{\non}{\nonumber}
\newcommand{\OO}{\mathcal{O}}

\newcommand{\DS}{\displaystyle}
\newcommand{\aes}{\left\langle}
\newcommand{\ad}{\right\rangle}
\newcommand{\al}{\alpha}
\newcommand{\bt}{\beta}
\newcommand{\gm}{\gamma}
\newcommand{\dt}{\delta}
\newcommand{\lb}{\lambda}
\def\skip1{\vskip \baselineskip}
\def\skip2{\vskip 2\baselineskip}
\def\dstl{\displaystyle}
\def\beq{\begin{equation}}
\def\eeq{\end{equation}}
\def\beqa{\begin{eqnarray}}
\def\eqa{\end{eqnarray}}
\begin{document}

%Page0 (Title Page)

\newcommand{\bm}[1]{\mbox{\boldmath$#1$}}

\thispagestyle{empty}

\tightenlines

\title{Effects of Random Biquadratic Couplings in a Spin-1 Spin-Glass Model}

\vskip \baselineskip

\author{J.M. de Ara\a'{u}jo,$^{1,2}$
F.A. da Costa,$^{2,3}$ and F.D. Nobre$^2$ }

\vskip \baselineskip

\address{
$^1$ Departamento de Ci\^{e}ncias Naturais \\
Universidade Estadual do Rio Grande do Norte \\
59610-210 \hspace{8mm} Mossor\'o - RN, \hspace{8mm} Brazil\\
$^2$ Departamento de F\a'{\i}sica
Te\a'{o}rica e Experimental \\
Universidade Federal do Rio Grande do Norte \\
Campus Universit\a'{a}rio -- \ C.P. 1641 \\
59072-970 \hspace{8mm} Natal - RN, \hspace{8mm} Brazil \\
$^3$ Present address: Instituto de F\a'{\i}sica \\
Universidade Federal Fluminense \\
Av. Litor\^anea s/n -- Boa Viagem \\
24210-340 \hspace{8mm} Niteroi - RJ, \hspace{8mm} Brazil }

\maketitle

\newpage

\begin{abstract}
%\vskip \baselineskip

A spin-$1$ model, appropriated to study the competition between
bilinear ($J_{ij}S_{i}S_{j}$) and biquadratic ($K_{ij}S_{i}^{2}S_{j}^{2}$)
random interactions, both of them with zero mean, is investigated. The
interactions are infinite-ranged and the replica method is employed. Within
the replica-symmetric assumption, the system presents two phases, namely,
paramagnetic and spin-glass, separated by a continuous transition line. The
stability analysis of the replica-symmetric solution yields, besides the
usual instability associated with the spin-glass ordering, a new phase due to
the random biquadratic couplings between the spins.

\end{abstract}

\vspace{5cm}

\begin{tabbing}
\=xxxxxxxxxxxxxxxxxx\= \kill
\>{\bf Keywords:} \> Spin Glasses; Biquadratic Couplings; Replica Method. \\
\>{\bf PACS Numbers:} \> 05.70.-a, 05.70.Fh, 64.60.-i \\
\end{tabbing}

\newpage

\section{Introduction}

%\noindent
The study of disordered systems has grown very fast during the last
years. Among these systems, spin glasses \cite{Young97B,Binder86,Fischer91}
have attracted much attention. One of its main caracteristic is the
existence of a very rugged free-energy landscape, with many minima separated
by high barriers. It turns out that the equilibrium state of
such system becomes hardly accessible in an experiment, as one may
guess. The spin-glass mean-field theory is well
established \cite{Binder86,Fischer91}, being highly nontrivial. However, the
effects of fluctuations around the mean-field solution are very difficult to
take into account in general cases, with most of the results been obtained
for the Edwards-Anderson model \cite{Dominicis97}.

Many other spin-glass models have been investigated within the mean-field
level (for reviews, see Refs. \cite{Young97B,Binder86,Fischer91} ). Recently,
much effort has been dedicated to understanding the phase behavior of
spin-1 Ising glasses \cite{XcYoNb97,daCosta94,Arenzon96,Arenzon97,Schreiber99},
as promising models to describe real systems which present multicritical
phenomena. Other models which can be mapped onto spin-1 Ising glasses were
also studied recently
\cite{Lutchinskaya84,Kopec93c,Kopec93b,LuTa95,Walasek95,Cesare96,Walasek97,Cesare97}.
However, as far as we know, none of those works addresses to the competition
between bilinear and biquadratic random interactions. In fact, a few years
ago a spin-glass version of the Blume-Emery-Griffiths model \cite{beg} was
introduced in order to describe disordered magnetic lattice gases
\cite{Inawashiro83,Inawashiro81,Inawashiro81b,Inawashiro82,Thompson86},
including both bilinear and biquadratic random couplings; however, due to
the fact that a replica stability analysis was not performed, an important
ingredient was missing, i.e., broken ergodicity, usually associated with
irreversibility effects. The purpose of the present work is to fill this
gap by investigating the overall behavior of a system which presents the
aforementioned random interactions in a simple spin-1 model. In order to
determine the free-energy density and the corresponding equations of
state, we will use the replica mean-field approach. Under the
replica-symmetry assumption \cite{SK75}, the system exhibits only
two phases separated by a continuous transition line. However, the
stability analysis of the replica-symmetric solution performed within the approach
proposed by de Almeida and Thouless \cite{AT78} suggests the existence
of three distinct phases. The paper is organized as follows. In Section
\ II we describe the model and obtain the replica free energy. The
replica-symmetric
solution, as well as the corresponding phase diagram is investigated in
Section \ III. The stability analysis of the replica-symmetric solution is performed in
Section \ IV. Our findings are summarized in Section \ V, where we also
present our conclusions.

\section{The Model and its Free-Energy Density }

In this paper we consider an infinite-range interaction spin-glass model
described by the Hamiltonian

\begin{equation}
H=-\sum_{(i,j)}J_{ij}S_{i}S_{j}-\sum_{(i,j)}K_{ij}S_{i}^{2}S_{j}^{2},
\label{Ham}
\end{equation}
where each spin $S_{i}$ \ ($i=1,2, \cdots , N$) can take the values
$-1,0$ and $1$ and the
summations are over all distinct pairs $(i,j)$. Both couplings are quenched,
independent random variables, following probability distributions

\begin{equation}
P(X_{ij})=\left( \frac{N}{2\pi X^{2}}\right)^{1/2} \exp \left(
-\frac{NX_{ij}^{2}}{2X^{2}}\right) ,  \label{gaussd}
\end{equation}
where $X$ stands for either $J$ or $K$. There are two obvious limiting cases
of this model. First, in the absence of biquadratic interactions
($K_{ij}=0$ for every pair $(i,j)$), we have a conventional spin-$1$
spin-glass model. The properties of this model are quite analogous to those
of the Sherrington-Kirkpatrick model
\cite{XcYoNb97,daCosta94,Arenzon96,Arenzon97,Schreiber99,Ghatak77,Lage82,Mottishaw85}; the system
presents a continuous transition from a paramagnetic to a low-temperature
spin-glass phase where the ergodicity is also broken \cite{Lage82}. On the other
hand, if $J_{ij}=0$ for every pair $(i,j)$, the system becomes equivalent to
the discrete quadrupolar-glass model investigated in Ref. \cite{Lutchinskaya84}. In
this case, there is no sharp transition to a low-temperature phase; however,
a stability analysis shows that in fact there is a phase transition to a
low-temperature nonergodic region \cite{Kopec93c}. We will be most interested
in the case where both $J_{ij}$ and $K_{ij}$ are distinct from zero, in order
to appreciate the effects of their competition on the phase diagram.

The free-energy density for this system is given by

\begin{equation}
\beta f=-\lim_{N\rightarrow \infty }\frac{\overline{\ln Z}}{N} , \label{fed1}
\end{equation}
where the bar denotes an average over the disorder. Such an average is
performed by the so-called replica method, through the identity

\begin{equation}
\ln Z=\lim_{n\rightarrow 0}\frac{Z^{n}-1}{n} ~, \label{logZ}
\end{equation}
which avoids the difficulty of averaging the logarithm. Using
standard procedures \cite{Binder86,Fischer91}, we obtain

\begin{equation}
\beta f=\lim_{n\rightarrow 0}\frac{1}{n}\min g_{n}(q_{\alpha \beta },Q_{\alpha
\beta },p_{\alpha }),  \label{fed2}
\end{equation}
where

\begin{equation}
g_{n}(q_{\alpha \beta },Q_{\alpha \beta },p_{\alpha })=\frac{1}{4}\sum_{\alpha
\neq \beta }\left( \beta ^{2}J^{2}q_{\alpha \beta }^{2}+\beta
^{2}K^{2}Q_{\alpha \beta }^{2}\right) +\frac{\beta ^{2}}{4}%
(J^{2}+K^{2})\sum_{\alpha }p_{\alpha }^{2}-\ln {\rm Tr}\exp ({\cal H}_{eff})
\label{fung}
\end{equation}
and

\begin{equation}
{\cal H}_{eff}=\frac{\beta ^{2}J^{2}}{2}\sum_{\alpha \neq \beta }q_{\alpha
\beta }S^{\alpha }S^{\beta }+\frac{\beta ^{2}K^{2}}{2}\sum_{\alpha \neq
\beta }Q_{\alpha \beta }\left( S^{\alpha }\right) ^{2}\left( S^{\beta
}\right) ^{2}+\frac{\beta ^{2}}{2}\left( J^{2}+K^{2}\right) \sum_{\alpha
}p_{\alpha }^{2}\left( S^{\alpha }\right) ^{2},  \label{Heff}
\end{equation}
with the indexes $\alpha $ and $\beta $ running from $1$ to $n$.
Stationarity of $g_{n}$ with
respect to $q_{\alpha \beta }$, $Q_{\alpha \beta }$ and $p_{\alpha }$ gives
the equations of state,

\begin{eqnarray}
p_{\alpha }& =\aes \left( S^{\alpha }\right) ^{2}\ad _{n},
\label{equil} \nonumber\\
q_{\alpha \beta }& =\aes S^{\alpha }S^{\beta }\ad _{n},
 \\
Q_{\alpha \beta }& =\aes (S^{\alpha }S^{\beta })^{2}\ad _{n},
\nonumber
\end{eqnarray}
where $\aes {~~}\ad _{n}$ denotes an average with respect to the ``effective
Hamiltonian'' in
Eq. (\ref{Heff}). Whereas the order parameters $q_{\alpha \beta }$ and
$Q_{\alpha \beta }$  are already expected, the free energy depends also on
$p_{\alpha}$, a disorder induced order parameter which measures the fraction
of spins in the states $S^{\alpha}=\pm 1$, for each replica $\alpha$.

In the following two sections we consider the replica-symmetric solution and
its corresponding stability analysis.

\section{Replica-Symmetric Solution }

The simplest solution of the saddle-point equations is the replica symmetric
Ansatz, which consists in assuming

\begin{eqnarray}
p_{\alpha }&=& p, \quad \forall \alpha  \nonumber  \\
q_{\alpha \beta }&=&q, \quad \forall  (\alpha  \beta)  \label{rsy}  \\
Q_{\alpha \beta }&=&Q, \quad \forall  (\alpha  \beta) .  \nonumber \
\end{eqnarray}

\noindent
Inserting this Ansatz into Eqs. (\ref{fed2})--(\ref{Heff})
and performing some simple Gaussian transformations, the free-energy
density becomes

\begin{equation}
f=\frac{\beta J^{2}}{4}\left( p^{2}-q^{2}\right) +\frac{\beta K^{2}}{4}%
\left( p^{2}-Q^{2}\right) - \aes\aes \ln z(x,y) \ad\ad_{xy}
\label{fed3}
\end{equation}
where

\begin{equation}
z(x,y)=1+2\exp (\Delta)\cosh \left( \beta J\sqrt{q}%
x\right) ,  \label{zxy}
\end{equation}

\begin{equation}
\Delta = \frac{\beta ^{2}J^{2}}{2}\left( p-q\right) +\frac{\beta
^{2}K^{2}}{2}\left( p-Q\right) +\beta K\sqrt{Q}y ,    \label{delxy}
\end{equation}
and

\beq
\aes\aes h(x,y)\ad\ad_{xy} =\dstl{\int_{-\infty }^{+\infty }\frac{\dstl dx}
{\dstl \sqrt{2\pi }}\int_{-\infty }^{+\infty }\frac{dy}{\sqrt{2\pi}}}
\exp(-\frac{x^2+y^2}{2})h(x,y) .
\label{aver}
\eeq
For the equations of state one gets,

\begin{eqnarray}
p& = \aes\aes \varphi_{2}(x,y)\ad\ad_{xy} , \label{eqp} \\
q& = \aes\aes \varphi_{1}^2(x,y) \ad\ad_{xy}, \label{eqq}\\
Q& = \aes\aes \varphi_{2}^2(x,y) \ad\ad_{xy} ,  \label{eqQ}\
\end{eqnarray}
with

\begin{eqnarray}
\varphi_{1}(x,y)&=\frac{\dstl 2e^{\Delta}}{\dstl z(x,y)}
                   \sinh(\beta J\sqrt{q}x) ~, \\
\varphi_{2}(x,y)&=\frac{\dstl 2e^{\Delta}}{\dstl z(x,y)}
                    \cosh(\beta J\sqrt{q}x) ~. \
\end{eqnarray}

As mentioned before, the $K=0$ case is equivalent to $D=0$ in the model
studied by Ghatak and Sherrington \cite{Ghatak77}. In this particular case
it is well known that we have a continuous transition from a paramagnetic
to a spin-glass phase at $k_{B}T/J=0.7901\cdots$. In general, the above
equations present a trivial solution $q=0$ but with $p\neq0$, $Q\neq0$
for any temperature. In the low-temperature regime we also find another
phase, with all order parameters distinct from zero. The phase boundary
separating these two phases can be easily obtained from an expansion in
powers of $q$, in either the free-energy density, or the equation of state
for $q$. In either way, we find a critical frontier given by

\beq
               Q = (k_{B}T/J)^2 . \label{crit}
\eeq
From the same expansions, we also ruled out the possibility
of first-order transitions and tricritical behavior. The condition given
by the above equation involves both order parameters $Q$ and $p$, which
should satisfy Eqs. (\ref{eqp}) and (\ref{eqQ}) with $q$ set to zero.
We are thus left with a set of three coupled nonlinear equations, which,
except for some particular limits, has no
analytical solution. We performed a detailed numerical study of these
equations in order to check for the possibility of other types of orderings,
but we found none, besides those already described. As a result of our
analysis, we found the critical frontier shown in Fig. \ref{diag}.
We performed an expansion for $K\gg J$, and verified
that assymptotically, such a critical frontier approaches the limit
$k_{B}T/J = 0.7876\cdots$, with $p \cong Q = 0.6204\cdots$. It is
important to mention that within the present analysis, the high-temperature
phase should be identified as an extension of the paramagnetic one
already present when $K = 0$. This can be justified by the following
argument: the free-energy density, as well as the order parameters $p$
and $Q$, may be expanded as power series of $K$, for small values of $K$.
Therefore, no anomalous behavior on the thermodynamical functions can be
seen as we let $K \rightarrow 0$. Similarly, the low-temperature phase should
be identified with the spin-glass phase ocurring at $K=0$. Thus, as far as
the replica symmetric solution is concerned, the biquadratic random coupling
does not bring any new physics to this system.

In the following section we will consider the stability of the
above-mentioned solutions against replica fluctuations, and it
will be shown that this study leads to an important modification on the
paramagnetic side.

\section{Stability of the replica-symmetric solutions }

Since the work of de Almeida and Thouless \cite{AT78}, it is generally
believed that replica-symmetric solutions are unstable under small fluctuations on
the whole replica space. In our case, these fluctuations are
governed by the Hessian matrix,

\beq
\label{hes}
{\bf G} =
\pmatrix{
{\DS\partial^2g_{n}\over\DS\partial p_\alpha\partial p_\beta} &
{\DS\partial^2g_{n}\over\DS\partial p_\alpha\partial q_{\nu\gamma}} &
{\DS\partial^2g_{n}\over\DS\partial p_\alpha\partial Q_{\nu\gamma}} \cr\cr
{\DS\partial^2g_{n}\over\DS\partial q_{\nu\gamma} \partial p_{\alpha}} &
{\DS\partial^2g_{n}\over\DS\partial q_{\alpha\beta}\partial q_{\nu\gamma}} &
{\DS\partial^2g_{n}\over\DS\partial q_{\alpha\beta}\partial Q_{\nu\gamma}} \cr\cr
{\DS\partial^2g_{n}\over\DS\partial Q_{\nu\gamma}\partial p_\alpha} &
{\DS\partial^2g_{n}\over\DS\partial Q_{\nu\gamma}\partial q_{\alpha\beta}} &
{\DS\partial^2g_{n}\over\DS\partial Q_{\alpha\beta}\partial Q_{\nu\gamma}}
}
\eeq

\vspace{28pt} \noindent
where $g_{n}$ is given by Eq. (\ref{fung}).  Stability requires that all
eigenvalues of this matrix, evaluated within the replica-symmetric solution, should
be positive (see Appendix A for the computation of such eigenvalues).
In the limit $n\rightarrow 0$ we get three
longitudinal eigenvalues, as the roots of the secular equation

\beq
\left |
\begin{array}{ccc}
{\cA}-{\BB}-\lambda ^{(L)} &  \DD-\CC            &    \FF-\EE          \\
2(\CC-\DD)      & {\GG}-4{\HH}+3{\II}-\lambda ^{(L)}  &  {\JJ}-4{\KK}+3{\LL} \\
2(\FF-\EE)      & {\JJ}-4{\KK}+3{\LL}    &  {\MM}-4{\NN}+3{\OO}-\lambda ^{(L)}
\end{array}
\right | =0 ~,  \label{long}
\eeq

\vspace{28pt} \noindent
and two transverse ones, given by

\beq
\lambda ^{(T)}={{\GG}-2{\HH}+{\II}+{\MM}-2{\NN}+{\OO} \pm
     \sqrt{[{\GG}-2{\HH}+{\II}-{\MM}+2{\NN}-{\OO}]^2 -
             4[{\JJ}-2{\KK}+{\LL}]^2}\over 2} \label{tev}
\eeq
where the quantities ${\cA},\ldots,{\OO}$ are defined in
Appendix A. In the paramagnetic phase, where $q=0$, one of the longitudinal
eigenvalues becomes

\beq
            \lambda _{1}^{(L)}=\left({J\over k_{B}T}\right)^2 -
                             \left({J\over k_{B}T}\right)^4Q ~,
\eeq
whereas the other two are given by

\beq
\lambda_{2,3}^{(L)}={{\cA}-{\BB}+{\MM}-4{\NN}+3{\OO} \pm
     \sqrt{[{\cA}-{\BB}-{\MM}+4{\NN}-3{\OO}]^2 -
             8({\EE}-{\FF})^2}\over 2} ~.
\eeq

Let us first consider the behavior of the above eigenvalues throughout the
paramagnetic phase. Using both analytical and numerical calculations,
we find that all three longitudinal eigenvalues are positive,
with $\lambda_{1}^{(L)}$ vanishing along the paramagnetic to spin-glass
transition line. However, considering the transverse eigenvalues
of Eq. (\ref{tev}), we notice that one of them, denoted by $\lambda_{1}^{(T)}$
(corresponding to the plus sign before the square root),
becomes identical to $\lambda_{1}^{(L)}$ everywhere in the paramagnetic phase,
including the critical frontier paramagnetic/spin-glass, where it also
vanishes. Besides, in the paramagnetic region the second transverse
eigenvalue is given by

\beq
          \lambda_{2}^{(T)} = {\MM}-2{\NN}+{\OO} ~.
\eeq
Our numerical analysis shows that as the temperature decreases, and for
high enough values of $K$, $\lambda_{2}^{(T)}$ becomes negative, throughout
the paramagnetic phase. This suggests an onset of irreversibility in the
paramagnetic phase, associated with an ergodicity breaking, as we cross
the line given by $\lambda_{2}^{(T)} = 0$. This effect is brought about by
fluctuations on the order parameter $Q_{\alpha \beta }$, which in turn was
generated by the random biquadratic couplings. We identify this region as a
new phase, which we will call biquadratic spin-glass phase, with
replica-symmetry breaking associated to the parameter $Q_{\alpha \beta }$;
this region should, then, be properly described by the Ansatz of Parisi
\cite{Mezard87a}. We have also found that the boundary
paramagnetic/biquadratic spin-glass (where $ \lambda_{2}^{(T)} = 0 $) is a
straight line with slope $\approx 0.077 $; such a numerical result is in
full agreement with the one found in Ref. \cite{Kopec93c}
($ k_{B}T/J' \approx 1.38$), if one considers the proper changes of spin
variables and summations in the Hamiltonian of Ref. \cite{Kopec93c}
(which leads to $K =18J'$).

We have also investigated numerically the behavior of all five
eigenvalues in the spin-glass phase. The transverse eigenvalue
$\lambda_{1}^{(T)}$ is negative through the whole
spin-glass phase; this means that irreversibility is also present
in this phase and so, a solution with replica-symmetry breaking should
be employed.

The phase diagram resulting from this analysis is shown in Fig.
\ref{diag2}. The paramagnetic to spin-glass as well as the paramagnetic
to biquadratic spin-glass phase boundaries should remain valid under a
Parisi-like treatment. However, it is possible that the biquadratic
spin-glass to spin-glass frontier changes under replica-symmetry breaking in
both matrices ${\bf Q}$ and ${\bf q}$. Thus, the corresponding boundary shown
in Fig. \ref{diag2}, which was obtained within the replica-symmetric solution,
should be seen as a rather schematic one, although there is no physical
reason to expect a substantial qualitative change. The correct treatment
based on Parisi's Ansatz is very difficult in this case, since it involves
nonlinear integro-differential equations at finite temperatures, which are
hard to solve numerically. Such an analysis is beyond the purpose of this
paper.

\section{Conclusions }

We have studied a solvable spin-1 model, including both bilinear and
biquadratic random exchanges, with zero means and variances $J$ and $K$,
respectively. The model was solved through the replica formalism. Three types
of order parameters were introduced to describe the system in the replica
space: a density ($p_{\alpha}$), which measures the fraction of spins in the
states $S^{\alpha}=\pm 1$ and two spin-glass-like matrices, represented by
the bilinear and biquadratic matrix elements $q_{\alpha\beta}$ and
$Q_{\alpha\beta}$, respectively.

The replica-symmetric solution leads to a continuous transition from a
high-temperature paramagnetic phase to a low-temperature spin-glass phase,
signaled by the onset of the spin-glass order parameter $q$. The corresponding
critical frontier is almost temperature-independent, especially for large
values of the variance $K$. We have also analysed the eigenvalues of the
stability matrix associated with fluctuations around the replica-symmetric
solutions. We verified numerically that one of the replicon eigenvalues is
always negative throughout the whole spin-glass phase, implying an
instability of the replica-symmetric solution. We have also noticed that
the paramagnetic phase presents a similar instability (associated with the
matrix elements $Q_{\alpha\beta}$), for sufficiently large values of the
variance $K$, giving rise to a new phase which we have called biquadratic
spin-glass phase. Such instabilities may be related to the
onset of irreversibility effects, i.e., the response functions could depend
on the history of the system (e.g., field-cooling and zero-field-cooling
measurements may lead to different results) \cite{Binder86,Fischer91}.
On the basis of our findings, many of them from numerical analysis, we
conclude that for $K\neq0$ the system presents at least three distinct phases,
in which two of them should be properly described through a replica-symmetry
breaking procedure. The frontier separating the biquadratic spin-glass and
spin-glass phases requires further investigation. It is not clear if a full
Parisi solution would change substantially its location. We hope to address
to this point in a future work.

%\vspace{.5in}
\newpage
\noindent
{\bf Acknowledgments} \\
One of us (FDN) thanks CNPq and Pronex/MCT (Brazil) for partial financial
support. FAC would like to thank FAPESP for partial finantial
support during his visit to the Instituto de F\'{\i}sica of the
Universidade de S\~{a}o Paulo under the grant {\#}97/14223-7. Finally,
all of us would like to thank the hospitality of the Centro Brasileiro
de Pesquisas F\'{\i}sicas.

\newpage

\begin{appendix}
\section{Stability analysis of the replica-symmetric solution}
The elements of the Hessian matrix {\bf G}, defined in Eq. (\ref{hes}),
are given by

\beqa
{\DS\partial^2g_{n}\over\DS\partial p_{\al}\partial p_{\bt}} &=
{1\over2}(\bt\kappa)^2\delta_{\al\bt} -
{1\over4}(\beta\kappa)^4\left[\aes(S^{\al}S^{\bt})^2\ad_{n}-
\aes(S^\al)^2\ad_{n} \aes(S^\bt)^2\ad_{n}\right] ,  \\
{\DS\partial^2g_{n}\over\DS\partial p_\al\partial q_{\bt\gm}} &=
- {1\over2}(\bt\kappa)^2(\bt J)^2 \left[\aes(S^\al)^2S^\bt S^\gm\ad_{n}
- \aes (S^\al)^2 \ad_{n} \aes S^\bt S^\gm \ad_{n} \right] , \\
{\DS\partial^2g_{n}\over\DS\partial p_\al\partial Q_{\bt\gm}} &=
- {1\over2}(\bt\kappa)^2(\bt K)^2 \left[\aes(S^\al S^\bt S^\gm)^2\ad_{n}
- \aes (S^\al)^2 \ad_{n} \aes (S^\bt S^\gm)^2 \ad_{n} \right] , \\
{\DS\partial^2g_{n}\over\DS\partial q_{\al\bt} \partial q_{\gm\dt}} &=
(\bt J)^2 \delta_{\al\bt} - (\bt J)^4 \left( \aes S^\al S^\bt S^\gm S^\dt \ad_{n}
- \aes S^\al S^\bt \ad_{n} \aes S^\gm S^\dt \ad_{n}\right) , \\
{\DS\partial^2g_{n}\over\DS\partial q_{\al\bt} \partial Q_{\gm\dt}} &=
- (\bt J)^2(\bt K)^2 \left[ \aes S^\al S^\bt (S^\gm S^\dt)^2 \ad_{n}
- \aes S^\al S^\bt \ad_{n} \aes (S^\gm S^\dt)^2 \ad_{n} \right] , \\
{\DS\partial^2g_{n}\over\DS\partial Q_{\al\bt} \partial Q_{\gm\dt}} &=
(\bt K)^2 \delta_{\al\bt} -
(\bt K)^4 \left[ \aes (S^\al S^\bt S^\gm S^\dt)^2 \ad_{n}
- \aes (S^\al S^\bt)^2 \ad_{n} \aes (S^\gm S^\dt)^2 \ad_{n}\right] , \
\eqa
where
\beq
           \kappa^2 = J^2 + K^2.
\eeq
For the replica-symmetric solution the eigenvectors of {\bf G} have the form

\beq
{\bf u} =
\pmatrix{
\epsilon_\al \cr
\eta_{\al\bt} \cr
\xi_{\al\bt} } ,
\eeq
where

\beq
\epsilon_\al = p_\al - p, \qquad
\eta_{\al\bt} = q_{\al\bt} - q, \qquad
\xi_{\al\bt} = Q_{\al\bt} - Q,
\eeq
represent Gaussian fluctuations. Following de Almeida and
Thouless \cite{AT78}, we start with the eigenvector totally
symmetric under replica-index permutations

\beq
     \epsilon_\al = a, \quad \eta_{\al\bt}=b, \quad \xi_{\al\bt} =c,
     \quad {\rm for} ~~\al,\bt = 1\ldots n ,
\eeq
which correspond to the longitudinal eigenvectors, according to the
conventional classification \cite{Dominicis85}. For a finite value of $n$, the
corresponding eigenvalues follow from

\beqa
\lb^{(L)} a &=& {\cA}a + (n-1){\BB}a + (n-1){\CC}b+{1\over 2}(n-2)(n-1){\DD}b
          \\
          && +(n-1){\EE}c+{1\over 2}(n-2)(n-1){\FF}c , \non\\
\lb^{(L)} b &=& 2{\CC}a+(n-2){\DD}a+{\GG}b+2(n-2){\HH}b+{(n-2)(n-3)\over 2}{\II}b
          \\
          && +{\JJ}c+2(n-2){\KK}c+{(n-2)(n-3)\over 2} {\LL}c , \non\\
\lb^{(L)} c &=& 2{\EE}a+(n-2){\FF}a+{\JJ}b+2(n-2){\KK}b+{(n-2)(n-3)\over 2}{\LL}b
          \\
          && +{\MM}c+2(n-2){\NN}c+{(n-2)(n-3)\over 2} {\OO}c ,  \non \
\eqa
where

\beqa
\cA &=& \left. {\DS\partial^2g_{n}\over\DS\partial p_{\al}\partial p_{\al}}
    \right| _{RS}
    = {(\bt\kappa)^2\over2}\left[ 1- {(\bt\kappa)^2\over2}(1-p)p\right] ,
\\
\BB &=& \left. {\DS\partial^2g_{n}\over\DS\partial p_{\al}\partial p_{\beta}}
      \right| _{RS}
      = {(\bt\kappa)^4\over4}\left( p^2-Q \right) ,
\\
\CC &=& \left. {\DS\partial^2g_{n}\over\DS\partial p_\al\partial q_{\al\bt}}
    \right| _{RS}
    = {(\bt J)^2(\bt\kappa)^2\over2}(p-1)q ,
\\
\DD &=& \left. {\DS\partial^2g_{n}\over\DS\partial p_\al\partial q_{\bt\gm}}
    \right| _{RS}
    = {(\bt J)^2(\bt\kappa)^2\over2}(pq - w) ,
\\
\EE &=& \left. {\DS\partial^2g_{n}\over\DS\partial p_\al\partial Q_{\al\bt}}
    \right| _{RS}
    = {(\bt\kappa)^2(\bt K)^2\over2}(p-1)Q ,
\\
\FF &=& \left. {\DS\partial^2g_{n}\over\DS\partial p_\al\partial Q_{\bt\gm}}
    \right| _{RS}
    = {(\bt\kappa)^2(\bt K)^2\over2}(pQ-W) ,
\\
\GG &=& \left. {\DS\partial^2g_{n}\over\DS\partial q_{\al\bt} \partial q_{\al\bt}}
    \right| _{RS}
    = (\bt J)^2\left[ 1 + (\bt J)^2(q^2-Q) \right] ,
\\
\HH &=& \left. {\DS\partial^2g_{n}\over\DS\partial q_{\al\bt} \partial q_{\al\gm}}
    \right| _{RS}
    = (\bt J)^4(q^2-w),
\\
\II &=& \left. {\DS\partial^2g_{n}\over\DS\partial q_{\al\bt} \partial q_{\gm\dt}}
    \right| _{RS}
    = (\bt J)^4(q^2-s),
\\
\JJ &=& \left. {\DS\partial^2g_{n}\over\DS\partial q_{\al\bt} \partial Q_{\al\bt}}
    \right| _{RS}
    = (\bt J)^2 (\bt K)^2(Q-1)q ,
\\
\KK &=& \left. {\DS\partial^2g_{n}\over\DS\partial q_{\al\bt} \partial Q_{\al\gm}}
     \right| _{RS}
     = (\bt J)^2 (\bt K)^2(Qq-w) ,
\\
\LL &=& \left. {\DS\partial^2g_{n}\over\DS\partial q_{\al\bt} \partial Q_{\gm\dt}}
     \right| _{RS}
     = (\bt J)^2 (\bt K)^2(Qq-v) ,
\\
\MM &=& \left. {\DS\partial^2g_{n}\over\DS\partial Q_{\al\bt} \partial Q_{\al\bt}}
    \right| _{RS}
    = (\bt K)^2 \left[1 + (\bt K)^2(Q-1)Q \right] ,
\\
\NN &=& \left. {\DS\partial^2g_{n}\over\DS\partial Q_{\al\bt} \partial Q_{\al\gm}}
    \right| _{RS}
    = (\bt K)^4 \left(Q^2 - W \right),
\\
\OO &=& \left. {\DS\partial^2g_{n}\over\DS\partial Q_{\al\bt} \partial Q_{\gm\dt}}
    \right| _{RS}
    = (\bt K)^4 \left(Q^2 - S \right) .
\
\eqa
In the above equations, different replica labels are distinct from one
another, whereas $..|_{RS}$ means elements of the Hessian matrix evaluated
within the replica-symmetric solution; the parameters $p$, $q$ and $Q$ are given
by Eqs. (\ref{eqp})-(\ref{eqQ}) and

\beqa
      S &=& \aes\aes \varphi_{2}^4 \ad\ad_{xy},  \\
      s &=& \aes\aes \varphi_{1}^4 \ad\ad_{xy} , \\
      v &=& \aes\aes \varphi_{1}^2\varphi_{2}^2 \ad\ad_{xy}, \\
      W &=& \aes\aes \varphi_{2}^3 \ad\ad_{xy},  \\
      w &=& \aes\aes \varphi_{1}^2\varphi_{2} \ad\ad_{xy}.  \
\eqa

The next step is to find the anomalous eigenvalues. These
correspond to breaking the symmetry of the vector ${\bf u}$ with
respect to one specific replica-index, denoted herein by $\theta$,

\beq
\left\{
\begin{array}{l}
 \epsilon^\alpha=a_1 \quad{\rm for}\quad \alpha=\theta \\
 \epsilon^\alpha=a_2 \quad{\rm for}\quad \alpha\ne\theta \\
\end{array}
\right .
\eeq
\beq
\left \{
\begin{array}{l}
 \eta^{\alpha\beta}=b_1 \quad{\rm for}\quad \alpha \,~{\rm or}~\, \beta =\theta \\
 \eta^{\alpha\beta}=b_2 \quad{\rm for}\quad \alpha,\beta \ne\theta
\end{array}
\right .
\eeq
\beq
\left \{
\begin{array}{l}
 \xi^{\alpha\beta}=c_1 \quad{\rm for}\quad \alpha \,~{\rm or}~\, \beta =\theta \\
 \xi^{\alpha\beta}=c_2 \quad{\rm for}\quad \alpha,\beta \ne\theta
\end{array}
\right .
\eeq

%\beqa
%\epsilon_\al = a_1, ~~\al=1\ldots n, ~~\al\neq\theta, \quad
%\epsilon_\theta = a_2 \non \\
%\eta_{\al\bt} = b_1, ~\al,\bt=1\ldots n,  ~~\al,\bt\neq\theta \quad
%\eta_{\al\theta} = b_2, \al=1\ldots n \\
%\xi_{\al\bt} = c_1, ~\al,\bt=1\ldots n,  ~~\al,\bt\neq\theta \quad
%\xi_{\al\theta} = c_2, \al=1\ldots n \non \
%\eqa
\noindent Orthogonality with the replica-symmetric eigenvector implies

\beq
a_2 = -{a_1\over n-1}, \quad b_2 = -{b_1\over n-2}, \quad c_2 =
-{c_1\over n-2}.
\eeq
Thus, the anomalous eigenvalues follow from

\beqa
\lambda^{(A)}a_1 &=& ({\cA}-{\BB})a_1 + (n-1)({\CC}-{\DD})b_1 +
(n-1)({\EE}-{\FF})c_1 , \\
\lambda^{(A)}b_1 &=& {n-2\over n-1}({\CC}-{\DD})a_1 + [{\GG} +
n{\HH} - (n-3){\II}]b_1
\non \\
                 && + [{\JJ}+n{\KK}-(n-3){\LL}]c_1 = 0 , \\
\lambda^{(A)}c_1 &=& {n-2\over n-1}({\EE}-{\FF})a_1 + [{\JJ} +
n{\KK} - (n-3){\LL}]b_1  \\
                 && + [{\MM}+n{\NN}-(n-3){\OO}]c_1 = 0 . \non \
\eqa
In the limit $n\rightarrow0$ the longitudinal and anomalous eigenvalues
coincide and may be obtained from Eq. (\ref{long}).

Finally, we must find the transverse eigenvalues. In this case, two
replica indices are fixed and the symmetry between replicas
is broken with respect to such indices (herein denoted by
$\theta$ and $\nu$). One has,

\beq
\left\{
\begin{array}{l}
 \epsilon^\alpha=a_3 \quad{\rm for}\quad \alpha=\theta \, ~{\rm or}~ \,
 \nu \\
 \epsilon^\alpha=a_4 \quad{\rm for}\quad \alpha\ne\theta,\nu \\
\end{array}
\right .
\eeq
\beq
\left \{
\begin{array}{l}
 \eta^{\alpha\beta}=b_3 \quad{\rm for}\quad \alpha \,~{\rm or}~\,
 \beta =\theta\,~{\rm or}~ \,\nu \\
 \eta^{\theta\alpha}=\eta^{\nu\alpha}=b_4 \quad{\rm for} \quad
 \alpha\ne\theta,\nu \\
 \eta^{\alpha\beta}=b_5 \quad{\rm for}\quad \alpha,\beta \ne\theta,\nu
\end{array}
\right .
\eeq
\beq
\left \{
\begin{array}{l}
 \xi^{\alpha\beta}=c_3 \quad{\rm for}\quad \alpha \,~{\rm or}\, \beta
 =\theta\,~{\rm or}~ \,\nu \\
 \xi^{\theta\alpha}=\xi^{\nu\alpha}=c_4 \quad{\rm for}\quad
 \alpha\ne\theta,\nu \\
 \xi^{\alpha\beta}=c_5 \quad{\rm for}\quad \alpha,\beta \ne\theta,\nu
\end{array}
\right .
\eeq
These eigenvectors must be orthogonal to both longitudinal and
anomalous ones. Thus, it follows that $a_3 = a_4 = 0$ and

\beqa
b_3 &= -(n-2)b_4 &= {1\over2}(n-2)(n-3)b_5, \non \\
c_3 &= -(n-2)c_4 &= {1\over2}(n-2)(n-3)c_5.  \
\eqa
From these observations, the transverse eigenvalues
are given by the secular equations

\beqa
\lb^{(T)}b_3 &=& ({\GG}-2{\HH}+{\II})b_3 + ({\JJ}-2{\KK}+{\LL})c_3, \\
\lb^{(T)}c_3 &=& ({\JJ}-2{\KK}+{\LL})b_3 + ({\MM}-2{\NN}+{\OO})c_3, \
\eqa
which are independent of $n$.

\end{appendix}

%\bibliographystyle{prsty}
%\bibliography{bib}

%%%%%%%%%%%%%%%%%%%%%%%%%%%%%%%%%%%%%%%%%%%%%%%%%%%%%%%%%%%%%%%%%%%%%
%%%%%%%%%%%%%%%       figures captions        %%%%%%%%%%%%%%%%%%%%%%%
%%%%%%%%%%%%%%%%%%%%%%%%%%%%%%%%%%%%%%%%%%%%%%%%%%%%%%%%%%%%%%%%%%%%%

\begin{figure}

%\centerline{\psfig{figure=diag2.ps,height=8cm,width=8cm}}

\caption{\it Phase diagram obtained within the replica-symmetric solution,
presenting a continuous transition from a high-temperature paramagnetic phase
(P) to a low-temperature spin-glass phase (SG).}

\label{diag}

\end{figure}

\vskip \baselineskip

\noindent
\begin{figure}

%\centerline{\psfig{figure=diag2.ps,height=8cm,width=8cm}}

\caption{\it Phase diagram resulting from the stability analysis of the
replica-symmetric solution. This solution is stable throughout the
paramagnetic (P) phase only. The biquadratic spin-glass phase (BSG), which
occurs for large values of $K$, is characterized by an instability of the
replica-symmetric paramagnetic solution. In the whole spin-glass phase (SG)
the replica-symmetric solution is unstable.}

\label{diag2}

\end{figure}

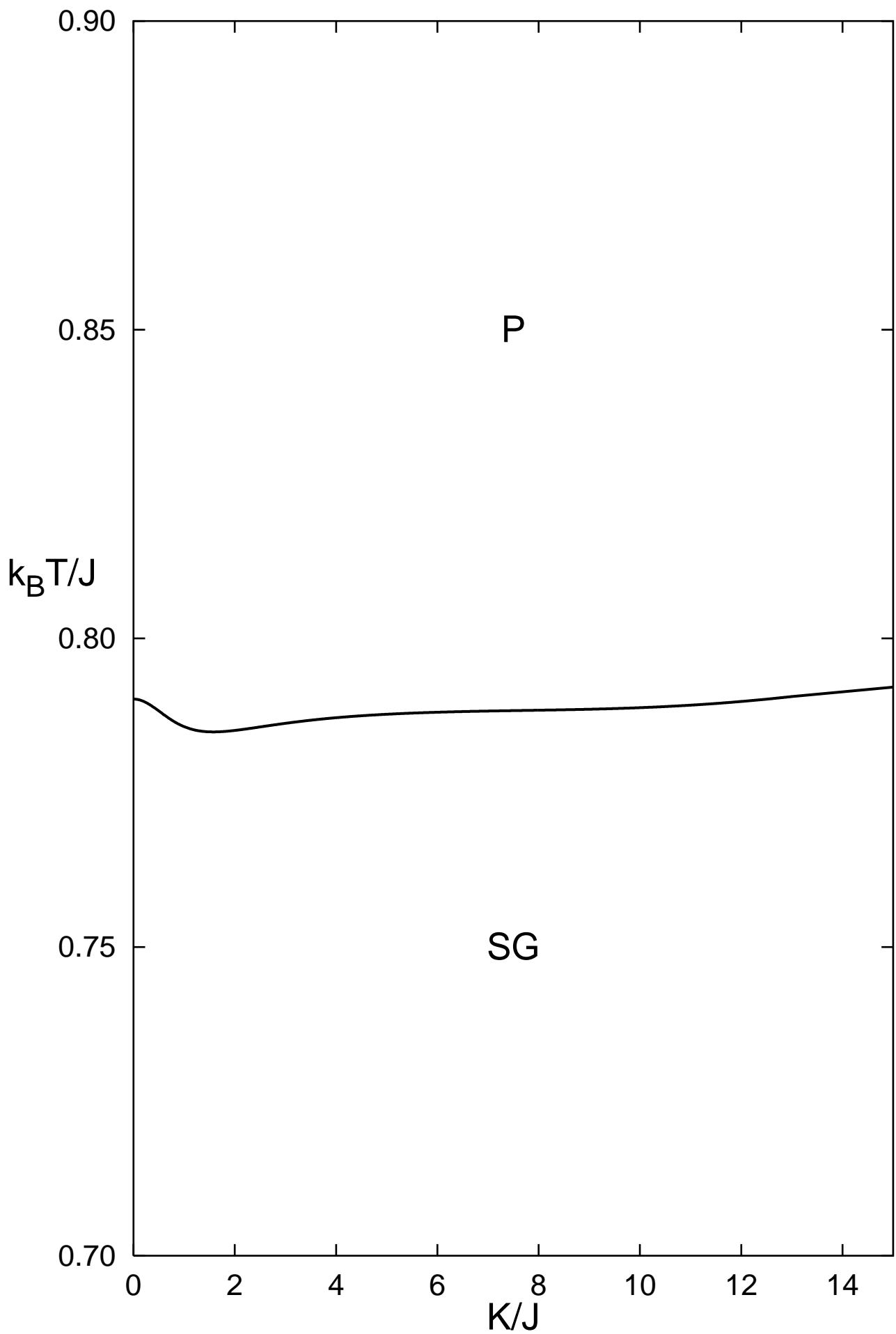
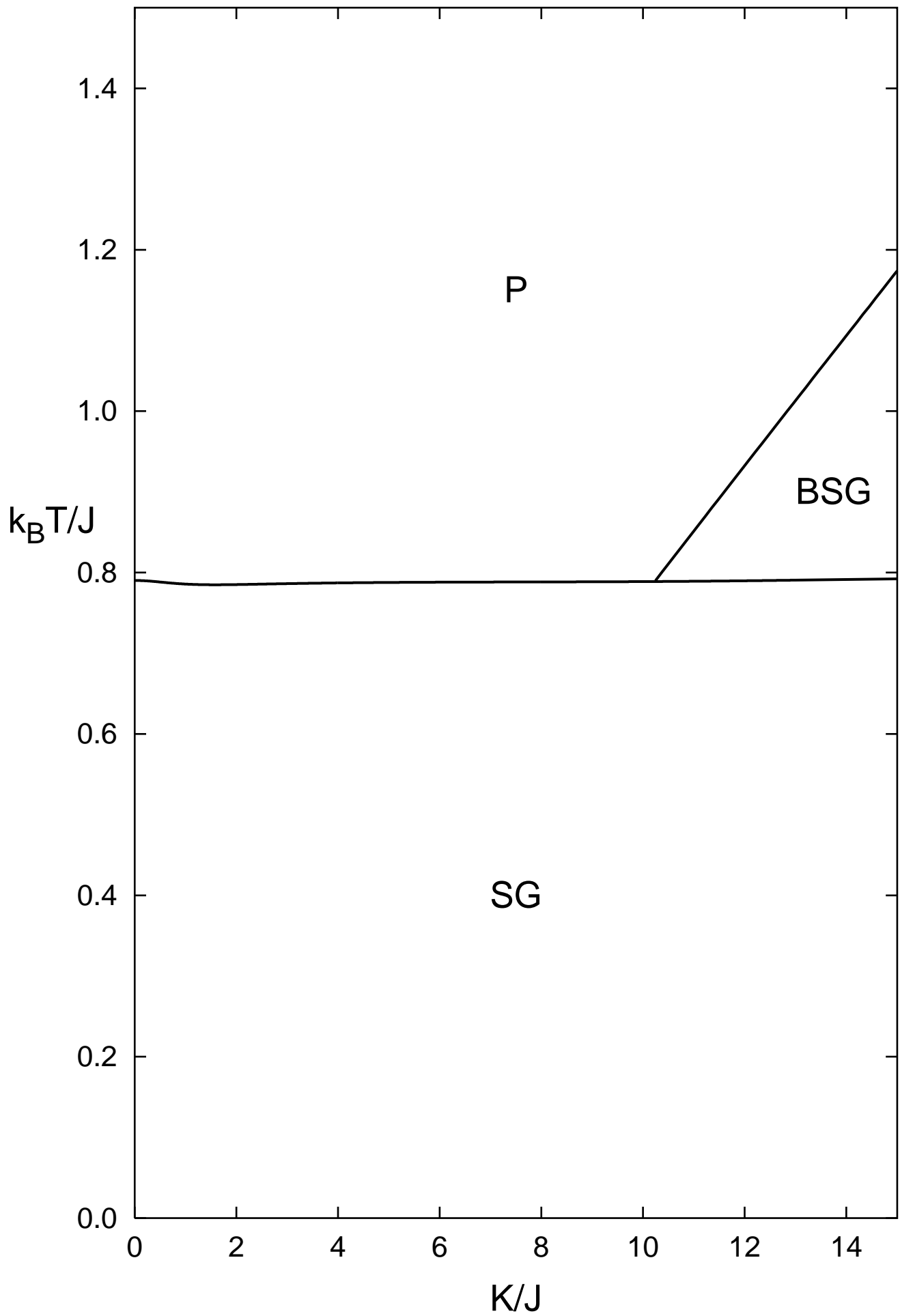
\end{document}